\documentstyle[12pt]{article}

\begin{document}
\hfill{UTTG-26-91}

\vspace{24pt}
\begin{center}
{\bf High Temperature Limit of the Confining Phase }

\vspace{24pt}
Joseph Polchinski

\vspace{12pt}
Theory Group\\ Department of Physics \\ University of Texas
\\ Austin, Texas 78712\\ bitnet:joe@utaphy

\vspace{12pt}
{\bf ABSTRACT}

\end{center}
\baselineskip=21pt

\begin{minipage}{4.8in}
The deconfining transition in non-Abelian gauge theory is known to occur
by a condensation of Wilson lines.
By expanding around an appropriate Wilson
line background, it is possible
at large $N$ to analytically continue the confining phase
to arbitrarily high temperatures, reaching
a weak coupling confinement regime.
This is used to study the high temperature
partition function of an $SU(N)$
electric flux tube.  It is found that
the partition function corresponds to
that of a string theory with a number of world-sheet fields that diverges
at short distance.
\end{minipage}

\vfill

\pagebreak

\setcounter{page}{1}

The large-$N$ limit is one of the few analytic methods available in strongly
coupled gauge theory, but has not yet proven solvable beyond two
dimensions.  Nevertheless, the
topological nature of the perturbation expansion
makes it tempting to believe that the large-$N$ theory can be
recast as a theory of noninteracting strings.$^1$
The principal tools that have been used to study this are the lattice
strong coupling expansion and
the loop equation.  The strong coupling expansion
can be written as a sum over noninteracting surfaces,
but the complicated weights make the continuum interpretation difficult.$^2$
The loop equations have
been argued to be equivalent to those for a string
theory with additional world-sheet fields,$^3$
but the mathematical intricacies
of the loop equations make this claim
difficult to evaluate.  In this paper we
will employ another tool, one which has proven useful in understanding the
physics of both gauge theory and string theory.  This is the study of the
vacuum amplitude in a periodic Euclidean spacetime.  We will focus on one
compact dimension of period $\beta$.  The vacuum amplitude can then be
interpreted as the partition function at inverse temperature $\beta$.

The basic question we wish to ask is the number of degrees of freedom of a
long flux tube at high energies.  In recent work with Strominger,$^4$ we
have investigated this question at low energies.  In this limit one
expects an effective string description with $D-2$
degrees of freedom, the transverse oscillations of the tube; here $D$ is the
spacetime dimension.  In ref.~4, this
was shown to be consistent with a covariant
conformal field theory of central charge~26, as needed for a consistent
quantization of the string.  The theory of ref.~4 is an effective theory,
nonrenormalizeable and so invalid at sufficiently high energies.  As one
increases the energy, one would not be surprised to find that the flux
tube has additional degrees of freedom, the
Liouville field$^5$ and the elf field$^3$
being possible candidates.  The number
of such fields at asymptotically high
energies, determined from the partition function,
would be an important piece of
information.  Further, since non-Abelian gauge
theory is asymptotically free, one
might hope that this could be obtained from a perturbative calculation.

There is a problem, pointed out by Thorn.$^6$  The high
temperature theory, in which one can calculate, is in a plasma phase and is
separated by a phase transition from the low-temperature confining phase in
which the flux tube exists.  The point of this paper is that at large $N$ it
is possible to analytically continue the low temperature phase past the
transition to arbitrarily high temperatures.  In
this way, one finds a regime in
which confinement and asymptotic freedom are simultaneously present.

To begin, we recall a remarkable parallel
between the statistical mechanics of
string theory and of large-$N$ gauge theory, as discussed
in detail by Atick and Witten.$^7$
In both theories the leading order free energy, $g_{\it string}^{-2}$
or $N^{-2}$ from the spherical
topology, is $\beta$-independent.  In string theory this occurs
because the spherical Riemann surface cannot wind around the
periodic dimension.  In gauge theory it occurs
because confinement restricts the
spectrum to color singlets, whose number goes as $N^0$.  In both theories
there is a transition at some $\beta_c$ to
a phase in which the leading order
free energy is $\beta$-dependent.

In $SU(N)$ gauge theory, where we need not initially
assume $N$ to be large, the
order parameters for this transition are the Wilson lines wrapping $k$ times
around the compact dimension $\tau$,$^8$
\begin{equation}
W_k(\vec x) = \frac{1}{N} {\rm Tr}\,P \exp \Bigl( i \int_{0}^{
k \beta} d\tau A^\tau(\tau, \vec x) \Bigr).
\end{equation}
The theory has
a $Z_{N}$ symmetry, consisting of
gauge transformations $g$ which are aperiodic by an element of the center,
\begin{equation}
g(\tau + \beta) = e^{2\pi i l /N} g(\tau).
\end{equation}
These multiply the order parameters by phases $e^{2\pi i kl /N}$.  At low
temperature, the Wilson line two-point function falls exponentially,
\begin{equation}
W_k (\vec R) W_{-k}(0) \sim e^{- \beta T_k (\beta) R},
\qquad R \to \infty.  \label{eq:twopoint}
\end{equation}
The two-point function vanishes at infinity, so the expectation values of the
order parameters are zero and the $Z_{N}$ symmetry is unbroken.  The Wilson
line $W_1$ is an external quark source, while the $W_k$ correspond to
combinations of sources in higher representations.  The $T_k(\beta)$ are
thus the effective string tensions for flux in the various
representations.

It is very
useful to consider also a dual point of view,
in which $\tau$ is regarded as a
spatial coordinate.  Then
\begin{equation}
M_k(\beta) = \beta T_k(\beta)
\end{equation}
are the masses of winding
states, states with electric flux in the periodic direction. At sufficiently
small $\beta$ these winding states become tachyonic (presumably the
singly-wound $M_1^2$ becomes negative first) and the theory makes a
transition to a phase in which the order
parameters have expectation values and
the $Z_{N}$ symmetry is broken.$^{8,9}$  These expectation
values mean that an
isolated quark source has finite energy: the theory no longer confines.

In the bosonic string theory there are also winding states, solitons with
vertex operators $e^{i k \beta (\tau_R - \tau_L)/ 4\pi }$ and masses
\begin{equation}
m^2_k(\beta) = -8\pi T + k^2 \beta^2 T^2, \label{eq:solmass}
\end{equation}
with $T = T_1(\infty)$ being the string tension.  These carry a
$U(1)$ symmetry, with world-sheet current $(\partial_z \tau,
 -\partial_{\bar z} \tau)$.$^{10}$
The phase structure is as in the gauge theory.
At large $\beta$ the $m^2$ are
all positive, and the $U(1)$ symmetry is unbroken.  At sufficiently small
$\beta$, $m_1^2$ becomes negative and the theory makes a transition
to a phase of broken $U(1)$ symmetry.$^{11,7}$

We are interested in the analytic continuation of the unbroken phase to small
$\beta$.  In string theory such a continuation is
quite routine.  Bosonic string
perturbation theory is an expansion around an unstable vacuum, due to
the $k=0$ tachyon in
eq.~(\ref{eq:solmass}).$^{12}$  In formal studies of the
perturbation theory this is usually an uninteresting artifact, and
amplitudes are defined by analytic continuation from
positive masses.  This can
be done for the $k \neq 0$ tachyons as well.
In particular, the tree-level
soliton masses-squared $m^2_k(\beta)$ given in eq.~(\ref{eq:solmass})
are simply analytic in
$\beta$.

Given the identical natures of the gauge and string transitions, the same
continuation should be possible in gauge theory.
We wish to expand the
high-temperature gauge theory around a non-standard  Wilson line background
with unbroken $Z_{N}$ (confinement), rather than the
broken symmetry phase.  Since the string continuation is carried out order
by order in $g_{\it string}$, we will make the corresponding expansion,
large-$N$, in the gauge theory.

To obtain an effective field theory of the Wilson lines,
fix the gauge such that the $\tau$ component of the vector potential
is $\tau$-independent and diagonal,
\begin{equation}
A_{ab}^\tau(\tau, \vec x) = \frac{\theta_a (\vec x)}{\beta} \delta_{ab},
\end{equation}
and carry out the one-loop integral over the spacelike components of the
vector potential.  This leaves an effective
three dimensional field theory of the
$\theta_a(\vec x)$,
\begin{equation}
\int [d \theta_a] e^{-S_{\it eff}}.
\end{equation}
with action$^{13}$
\begin{equation}
S_{\rm eff} = \int d^3 x\, \Biggl\{
\frac{1}{2 g^2 \beta} \vec \nabla \theta_a \cdot \vec \nabla \theta_a
+ \frac{1}{24\pi^2 \beta^3}
\sum_{a,b=1}^N \theta_{ab}^2 (2\pi - \theta_{ab})^2
\Biggr\},
\end{equation}
where $\theta_{ab}$ is $\theta_a - \theta_b$ modulo $2\pi$, defined to lie
between $0$ and $2\pi$.
The gradient term is from the tree level action and the potential term is
a one loop effect.  In the regime we will
consider, the symmetric phase at small
$g$, this is an honest low energy effective
theory: the $\theta_a$ have masses
of order $(g^2 N)^{1/2}\beta^{-1}$,
while the spacelike gauge bosons which have been
integrated out have masses of order $\beta^{-1}$.$^{14}$
There are corrections
of higher order in $g^2$, and also corrections of higher order in
$\beta^2 \nabla^2$.  Since the one
loop term arises from the scale $\beta$,
the coupling in the action should be understood as $g(\beta)$.

In the large-$N$ limit, the $\theta_a$ become infinite in number and it
becomes appropriate to consider the normalized density$^{15}$
\begin{equation}
\rho(\theta,\vec x) = \frac{1}{N}\sum_a \delta(\theta - \theta_a(\vec x)).
\end{equation}
The action becomes
\begin{eqnarray}
S_c &=& N^2 \int d^3 x\, \int_0^{2\pi} d\theta\,
\frac{1}{2 g^2
N \beta \rho} (\partial_\theta^{-1} \vec\nabla \rho)^2 \\[2pt]
&&\qquad +  \frac{N^2}{24\pi^2 \beta^3} \int d^3 x\,
\int \int_0^{2\pi} d\theta_1 d\theta_2 \,
\rho(\theta_1) \rho(\theta_2) \,\theta_{12}^2 (2\pi - \theta_{12})^2.
\nonumber
\end{eqnarray}
As usual in the large-$N$ limit, this is of order $N^2$ with $g^2 N$ held
fixed.  The potential favors all eigenvalues being equal,
\begin{equation}
\rho_b(\theta,\vec x) = \delta(\theta - \theta_0),
\end{equation}
breaking the $U(1)$ of $\theta \to \theta + \epsilon$.  The
symmetric phase is completely determined by
the requirement that the eigenvalue
distribution be $U(1)$-invariant,
\begin{equation}
\rho_s(\theta,\vec x) = \frac{1}{2\pi}.
\end{equation}

Expanding around the symmetric point,
\begin{equation}
\rho(\theta,\vec x) = \frac{1}{2\pi} \Biggl\{
1 + \sum_{ {k = -\infty} \atop {k\neq 0} }^\infty \rho_k(\vec x) e^{-ik
\theta} \Biggr\} ,
\end{equation}
the quadratic part of the action becomes
\begin{equation}
S_c^{(2)} = N^2 \sum_{ {k = -\infty} \atop {k\neq 0} }^\infty
\int d^3 x\, \Biggl\{ \frac{1}{2 g^2 N \beta k^2} (\vec\nabla \rho_k)^2
-\frac{1}{\pi^2 \beta^3 k^4} \rho_k^2 \Biggr\}.
\end{equation}
The Wilson line is simply $W_k(\vec x) = \rho_k(\vec x)$ so we can read off
the winding state masses
\begin{equation}
M_k^2 = -\frac{2 g^2 N}{\pi^2 \beta^2 k^2}. \label{eq:qcdmasses}
\end{equation}
This pattern is quite a bit different from that of the critical string,
eq.~(\ref{eq:solmass}).  As a check,
both spectra depend on the parameters $k$ and
$\beta$ in the combination $k \beta$.$^{16}$  This is because a {\it planar}
surface winding $k$ times around a dimension of period $\beta$ is the same
as one winding once around a dimension of period $k\beta$.  We can therefore
restrict attention to $k=1$.

It is convenient that the quantity the we are interested in,
the number of degrees of freedom of a long $k=1$ string, is
directly related by a modular transformation to
$M^2_1(\beta)$.$^{17}$
Suppose that the density of states of the flux tube
is equal to that of $n$ free world-sheet scalars, so the spectrum of a tube
with ends fixed at $0$ and $\vec R$ is
\begin{equation}
E^2 = T^2 R^2 + 2\pi T \sum_{a = 1}^{n} \Biggl( -\frac{1}{24} +
\sum_{m = 1}^{\infty} m N_{am} \Biggr),
\end{equation}
where $T$ is the string tension, $a$ labels the $n$ scalars,
$m$ labels the harmonic, and $N_{am}$ is the
occupation number.  We have
included the Casimir
constant $-\frac{1}{24}$ for good form, although it does
not matter much at high temperature.
The partition function Tr$(e^{-\beta E})$
can be found by standard methods.$^{18}$  It is
of the form~(\ref{eq:twopoint}) with
\begin{equation}
\beta^2 T_1^2(\beta) \equiv M_1^2(\beta) =  -\frac{n \pi T}{3} + \beta^2
T^2.  \label{eq:dof}
\end{equation}
At sufficiently small $\beta$ this is always negative, the entropy of
fluctuations overwhelming the string tension and favoring long strings.
This is the usual picture of the Hagedorn transition, equivalent to the
symmetry-breaking picture presented earlier.$^{11,7}$

Comparing equations~(\ref{eq:qcdmasses})
and~(\ref{eq:dof}), we see that the partition
function of the flux tube is not consistent with any fixed number of degrees
of freedom.  Rather, as we go to smaller $\beta$ we excite more and more
degrees of freedom, with
\begin{equation}
n_{\it eff}(\beta) \sim \frac{6 g^2(\beta) N }
{\pi^3 T \beta^2}. \label{eq:ndof}
\end{equation}
This is our main result.
Any string theory equivalent to large-$N$ gauge theory will have a
number of fields which diverges at short distances.$^{19}$  In particular,
various proposals that the gauge theory might be
equivalent to a string theory with a finite number of world-sheet fields
would seem to be ruled out.  This will extend to the spectra of mesons and
glueballs as well: the number of states at high energy will greatly
exceed that in a string theory.

This is complementary to the result obtained by Atick and Witten in a very
similar way.$^{7}$  They show that at high energy string theory has fewer
degrees of freedom than any field theory.
It should then be no surprise that
the large-$N$ gauge theory, which
is a field theory with moreover an infinite
number of fields, should correspond to
a string theory with an unusually large
number of degrees of freedom.
It is remarkable that we have been able to
use perturbation theory to obtain the number of degrees of freedom of an
object which exists only nonperturbatively.  We
have little physical intuition
for how this is possible, but experience with string theory gives us
confidence that the approach is correct.  The high temperature continuation
of the confining phase is quite similar to the Eguchi-Kawai
reduction of the lattice theory,$^{20}$ as we will discuss further below.

Now let us make some general remarks about the program to reformulate
the large-$N$ theory as a string
theory.  There are at least three ways in which
the flux tube might be unlike a fundamental string.  First, it might have
contact interactions, such as
self-avoidance, which would be nonlocal on the
world-sheet.  The large-$N$ factorization
property$^{21}$ shows that this is
not the case, since factorization holds even for sources in close proximity.
This also follows at least formally from
the strong-coupling sum over surfaces
representation.$^2$

Second, the world-sheet might have holes: the large-$N$ perturbation theory
looks like a net rather than a surface.  Here again the large-$N$
flux tube is indistinguishable from a
fundamental string:$^7$ as already noted the spherical amplitude
in a spacetime of large period $\beta$ is $\beta$-independent in both
theories, indicating no holes, and in both theories holes
develop spontaneously at small
$\beta$ from the condensation
of winding states.

Third, the flux tube might be fat,
unlike the infinitely thin fundamental string.
Indeed, the number of degrees of freedom~(\ref{eq:ndof}) grows as
$\beta^{-2}$, aside from the slow
running of the coupling.  This is just what
would be seen if the flux tube were a {\it three}-dimensional object with a
thickness of order $T^{-1/2}$.  Nevertheless, we believe that these degrees
of freedom are actually internal, and that the large-$N$ string has in some
sense a zero intrinsic thickness.  The first argument for this is
the formal lattice representation as a sum over surfaces.$^2$
The second comes from another application
of the high-temperature continuation.
Consider a large Wilson loop lying in a plane, and compactify the two
dimensions orthogonal to the plane.  If the flux tube had an intrinsic
thickness, we would expect that reducing the periods in the orthogonal
directions would squeeze the tube and increase its energy,
reducing the Wilson loop expectation value.  In fact, the expectation
value is independent of the size of the orthogonal directions at large $N$,
because a planar surface cannot wind around
these dimensions in the symmetric
phase.$^{22}$

We have used the high temperature continuation to count
the number of degrees
of freedom in the flux tube and to argue that the tube has zero thickness in
the large-$N$ limit.  Thus is should be equivalent to a {\it two-dimensional}
field theory with an infinite number of fields.  In fact, the high
temperature continuation makes it possible to construct this theory:
compactify
the orthogonal dimensions to very small radii while remaining in the
symmetric phase.  The resulting theory is essentially the dimensional
reduction of the gauge theory to two dimensions.  The degrees of freedom
are the $N^2$ gauge fields $A^\mu(\tau,\sigma)$; the main subtlety seems to
be in the expansion around the symmetric point.  Going further, compactifying
all but one dimension yields a matrix quantum mechanics of $A^\mu(\tau)$
whose spectrum is equivalent to the original gauge theory.
This result is already known from the Eguchi-Kawai
reduction,$^{20,23}$ which we see can be interpreted as the analytic
continuation of the confining phase.
This reduction has not yet proven analytically tractable,
but perhaps a reexamination is in order.

\centerline{\bf Acknowledgements}

I would like to thank L. McLerran, D. Minic, A. Strominger, and Z. Yang
for helpful dicussions.  Much of this work was carried out while the
author was a guest at the Aspen Center for Physics.
This research was supported in part by the
Robert A. Welch Foundation and NSF Grant PHY 9009850.

\vfill

\pagebreak

\centerline{\bf References}

\begin{itemize}
\item[1.] G. 't Hooft, Nucl. Phys. {\bf B72}, 461 (1974); {\bf B75}, 461
(1974).
\item[2.] V. A. Kazakov, Phys. Lett. {\bf B128}, 316 (1983);
V. I. Kostov, Phys. Lett. {\bf B138}, 191 (1984);
K. H. O'Brien and J.-B. Zuber, Nucl. Phys. {\bf B253}, 621 (1985).
\item[3.] For a review see A. A. Migdal, Phys. Rep. {\bf 102}, 199 (1983).
\item[4.] J. Polchinski and A. Strominger, ``Effective String Theory,''
Texas preprint UTTG-17-91 (1991).
\item[5.] A. M. Polyakov, Phys. Lett. {\bf B103}, 207 (1981).
\item[6.] C. Thorn, Phys. Lett. {\bf B99}, 458 (1981);
R. D. Pisarski, Phys. Rev. {\bf D29}, 1222 (1984).
\item[7.] J. Atick and E. Witten, Nucl. Phys. {\bf B310}, 291 (1981).
\item[8.] A. M. Polyakov, Phys. Lett. {\bf B72}, 447 (1978);
L. Susskind, Phys. Rev. {\bf D20}, 2610 (1979);
for a review see B. Svetitsky, Phys. Rep. {\bf 132}, 1 (1986).
\item[9.] It is of some interest whether the transition is first order,
occuring before $T_1(\beta)$ actually reaches zero, but this will not be
relevant for us.
\item[10.] This is the $U(1)$ which couples to
winding number.  Both the gauge
theory and the string theory have the additional $U(1)$ symmetry of
$\tau$-translations.  This is unbroken in both phases.
\item[11.] B Sathiapalan, Phys. Rev. {\bf D35}, 3277 (1987);
Ya. I. Kogan, JETP Lett. {\bf 45}, 709 (1987);
K. H. O'Brien and C.-I. Tan, Phys. Rev. {\bf D36}, 1184 (1987).
\item[12.] The $k=0$ tachyon is neutral under the $U(1)$.  It is
not clear whether it has any analog in the gauge theory.
\item[13.] D. J. Gross, R. D. Pisarski, and L. G. Yaffe, Rev. Mod. Phys.
{\bf 53}, 43 (1981).  See also
N. Weiss, Phys. Rev. {\bf D24}, 475 (1981);
{\bf D25}, 2667 (1982).
\item[14.] This is in contrast to
the unconfined plasma phase, where the spacetime vector potential has a
magnetic mass~$g^2 N \beta^{-1}$.
\item[15.] E. Br\'ezin, C. Itzykson, G. Parisi, and J.-B. Zuber, Comm.
Math. Phys. {\bf 59}, 35 (1978);
A. Jevicki and B. Sakita, Nucl. Phys. {\bf B165}, 511 (1980).
\item[16.] In particular, a careful treatment of the loop corrections would
show that the coupling $g$ in eq.~(\ref{eq:qcdmasses}) must be
evaluated at the scale
$k\beta$.
\item[17.] This sort of relation between ground state energy and asymptotic
degrees of freedom is well-known in string theory.  See, for example,
J. Cardy, Nucl. Phys. {\bf B270 [FS16]}, 186 (1986).  For a recent
discussion, see D. Kutasov and N. Seiberg, Nucl. Phys. {\bf B358}, 600 (1991).
\item[18.] R. Hagedorn, Nuovo Cim. Suppl. {\bf 3}, 147 (1965).
To derive eq.~(\ref{eq:dof}), one can use the integral
\begin{eqnarray*}
e^{\beta E} = \frac{\beta}{2\pi} \int_0^\infty \frac{ds}{s^{3/2}}
e^{(E^2 s + \beta^2 s^{-1})/2}
\end{eqnarray*}
to remove the square root from the exponent, carry out the sum over
occupation numbers, use the asymptotic formula
\begin{eqnarray*}
\prod_{m=1}^\infty (1 - x^m) \stackrel{x \to 1^-}{\sim} e^{-\pi^2/ 6\ln x},
\end{eqnarray*}
and then use the above integral again.
\item[19.] It may be unfamiliar that
the number of degrees of freedom can vary
with world-sheet scale, since the world-sheet
theory is conformally invariant.
The point is that in the noncritical string
the conformal invariance is broken
by the expectation value of the metric, $e^\phi$ or $\partial_z X^\mu
\partial_{\bar z} X_\mu$, making massive fields possible.
\item[20.] T. Eguchi and H. Kawai, Phys. Rev. Lett. {\bf 48}, 1063 (1982).
See ref.~3 for a review.
\item[21.] E. Witten, in NATO Advanced Studies Institute
Series---Series B: Physics, Vol. 59, eds. G. 't Hooft, et. al.
(Plenum, 1979).
\item[22.] This property of the confining phase
was derived from the lattice Schwinger-Dyson equations by
A. Gocksch and F. Neri, Phys. Rev. Lett. {\bf B50}, 1099 (1983).
\item[23.] A continuum version of the Eguchi-Kawai reduction was discussed
in D. J. Gross and Y. Kitazawa, Nucl. Phys. {\bf B206}, 440 (1982) and in
A. A. Migdal, Phys. Lett. {\bf B116}, 425 (1982).
\end{itemize}

\end{document}